\documentclass[aps,prl,reprint,superscriptaddress]{revtex4-2}

\newcount\showkey \showkey=0  

\newcount\draft \draft=1

\usepackage{color}
\usepackage{tabularx}

\usepackage{hyperref}

\usepackage{empheq}

\ifnum\showkey=1
\usepackage{seqsplit}
\usepackage[color]{showkeys}
\definecolor{refkey}{rgb}{0.9, 0.43, 0.63}
\definecolor{labelkey}{rgb}{0.59, 0.43, 0.63}
\usepackage{xstring}
\renewcommand*\showkeyslabelformat[1]{%
\noexpandarg%
\StrSubstitute{\(\{\)#1\(\}\)}{ }{\textvisiblespace}[\TEMP]%
\parbox[t]{\marginparwidth}{\raggedright\normalfont\small\ttfamily\expandafter\seqsplit\expandafter{\TEMP}}}
\fi

\usepackage{amsmath}
\usepackage{amssymb}
\usepackage{graphicx,multirow,subfigure}
\usepackage{float} 
\usepackage[T1]{fontenc}

\usepackage{rotating} 

\usepackage{enumitem}
\setlist[enumerate,2]{leftmargin=0.45em}

\usepackage{comment}

\usepackage[dvipsnames]{xcolor}

\newcommand{\nn}{\nonumber}

\renewcommand{\bar}{\overline}

\usepackage{physics}
\usepackage{bbold}
\usepackage{mathtools}

\renewcommand \ket[1]{
        \left| #1 \right>
}

\renewcommand \bra[1]{
        \left< #1 \right|
}

\newcommand{\beq}{\begin{equation}}
\newcommand{\eeq}{\end{equation}}
\newcommand{\beqa}{\begin{eqnarray}}
\newcommand{\eeqa}{\end{eqnarray}}
\newcommand{\bea}{\begin{eqnarray}}
\newcommand{\eea}{\end{eqnarray}}
\newcommand{\bi}{\begin{itemize}}
\newcommand{\ei}{\end{itemize}}
\newcommand{\ben}{\begin{enumerate}}
\newcommand{\een}{\end{enumerate}}
\newcounter{mycount}
\newcommand{\pauseen}{\setcounter{mycount}{\value{enumi}}\end{enumerate}}
\newcommand{\resumeen}{\begin{enumerate}\setcounter{enumi}{\value{mycount}}}

\newcounter{defcount}
\newcounter{myenum}

\newcommand\Tstrut{\rule{0pt}{2.6ex}}         

\begin{document}

\title{Determination of the $D\rightarrow \pi\pi$ ratio of penguin over tree diagrams}

\author{Margarita Gavrilova}
\email{mg2333@cornell.edu}
\affiliation{Department of Physics, LEPP, Cornell University, Ithaca, NY 14853, USA}
\affiliation{Kavli Institute for Theoretical Physics, University of California, Santa Barbara, CA 93106, USA}

\author{Yuval Grossman}
\email{yg73@cornell.edu}
\affiliation{Department of Physics, LEPP, Cornell University, Ithaca, NY 14853, USA}

\author{Stefan Schacht}
\email{stefan.schacht@manchester.ac.uk}
\affiliation{Department of Physics and Astronomy, University of Manchester, Manchester M13 9PL, United Kingdom}

\begin{abstract}
We study  the penguin over tree ratio in $D\rightarrow \pi\pi$ decays. This ratio can serve as a probe for rescattering effects. Assuming the Standard Model and in the isospin limit, we derive expressions that relate both the magnitude and the phase of this ratio to direct CP asymmetries and branching fractions. We find that the current data suggest that rescattering is large. A dedicated experimental analysis with current and future data will be able to significantly reduce the errors on these determinations, and enable us to check if indeed there is significant rescattering in $D \to \pi \pi$ decays.
\end{abstract}

\maketitle

\section{Introduction \label{sec:introduction}}
After the discovery of charm CP violation through the measurement of the difference of CP asymmetries of singly-Cabibbo suppressed (SCS) decays~\cite{LHCb:2019hro}, the 
recent first evidence of CP violation in a single decay channel, $D^0\rightarrow \pi^+\pi^-$~\cite{LHCb:2022lry}, showed hints for an enhanced $\Delta U=1$ contribution to the subleading amplitude~\cite{Schacht:2022kuj}. These new developments make searches for charm CP violation in additional channels 
particularly interesting in order to probe the effects of non-perturbative QCD and physics beyond the Standard Model (BSM)~\cite{Bause:2022jes}.

CP violation in SCS charm decays has been studied for a long time, especially in the context of U-spin and SU(3)$_F$ symmetry~\cite{Golden:1989qx, Buccella:1994nf, Pirtskhalava:2011va, Bhattacharya:2012ah, Hiller:2012xm, Cheng:2012wr, Feldmann:2012js, Grossman:2012ry, Nierste:2015zra,  Muller:2015lua, Muller:2015rna, Nierste:2017cua, Grossman:2018ptn, Cheng:2019ggx}. 
SU(3)$_F$ breaking effects have also been modelled accounting for 
final-state interactions~\cite{Buccella:2013tya, Buccella:2019kpn}
or in the context of the factorization-assisted topological-amplitude approach~\cite{Li:2012cfa, Li:2019hho}.
Flavor-symmetry methods have been applied to $D$~mixing, too, see Refs.~\cite{Falk:2001hx, Falk:2004wg, Kagan:2020vri}. Note that in this case 
a treatment within the heavy quark expansion is consistent with the data due to large theory uncertainties~\cite{Lenz:2020efu}.

In this letter, we study CP violation in $D \rightarrow \pi\pi$ decays. The decay amplitudes of $D \rightarrow \pi\pi$ decays can be written as  
\begin{align}
    A^f &= -\lambda_{d} A_{d}^f - \frac{\lambda_b}{2}A_b^f\,, \label{eq:SCS-amps}
\end{align}
where $f$ denotes the respective final state and $\lambda_{d,b}$ are the combinations of Cabibbo-Kobayashi-Maskawa~(CKM) matrix elements
\begin{equation}\label{eq:lambda_q}
    \lambda_{d} = V_{cd}^* V_{ud}\,, \quad \lambda_b = V^*_{cb} V_{ub}.
\end{equation}
$A_d^f$ and $A_b^f$ are the CKM-leading and -subleading amplitudes, respectively. In the parametrization of decay amplitudes in Eq.~(\ref{eq:SCS-amps}) we use the convention that $A_d^f$ and $A_b^f$ contain strong phases only.

The key quantity we focus on in this letter is the magnitude of the ratio of CKM-subleading over CKM-leading amplitudes
\begin{align}\label{eq:rf-def-1}
r^f &\equiv \vert A_b^f/A_d^f\vert\,.
\end{align}
This ratio is also known as \lq\lq{}penguin over tree ratio\rq\rq{}.
Note that the short distance contributions to the amplitudes $A^f_{b}$ are very small due to the very effective GIM suppression and we neglect them throughout this letter. The CKM-subleading amplitude $A_b^f$, to which we refer as penguin, thus contains only the long distance effects, also known as rescattering.

As we explain in more detail below, in the limit of no final state rescattering $r^f = 0$. Thus, a measurement of $r^f$ can serve as a probe of rescattering effects in $D \rightarrow \pi\pi$ decays.

The question of the size of the rescattering is a hint to a deeper question about QCD: can we treat QCD as perturbative at the charm scale? If it is perturbative, rescattering should be small, that is $r^f \ll 1$. Non-perturbative effects will result in $r^f \gtrsim 1$. Thus, a clear way to probe that ratio is called for.

Although there are some first conceptual ideas on the lattice~\cite{Hansen:2012tf} there is at this time no reliable way to calculate $r^f$. The available theory estimates for this ratio vary depending on the employed methodology.
Light cone sum rule (LCSR) calculations~\cite{Khodjamirian:2017zdu, Chala:2019fdb} find this ratio to be $\mathcal{O}(0.1)$. 
Estimates allowing for larger non-perturbative effects such as large rescattering or nearby resonances result in $r^f \sim\mathcal{O}(1)$~\cite{Soni:2019xko, Schacht:2021jaz, Brod:2011re, Grossman:2019xcj}, see also Refs.~\cite{Golden:1989qx, Brod:2012ud}. Studies employing coupled dispersion relations and rescattering data can be found in Ref.~\cite{Bediaga:2022sxw,Pich:2023kim, Franco:2012ck}.
While Ref.~\cite{Bediaga:2022sxw} explains the charm CP violation data within the Standard Model (SM), Ref.~\cite{Pich:2023kim} finds that this not possible. Ref.~\cite{Franco:2012ck} finds enhanced charm CP violation to be marginally consistent with rescattering effects.

Although the  estimates for $r^f$ come with a large theory uncertainty, its sensitivity to both rescattering effects and BSM physics~\cite{Grossman:2006jg, Altmannshofer:2012ur, Dery:2019ysp, Bause:2022jes, Calibbi:2019bay, Bause:2020obd, Buras:2021rdg, Acaroglu:2021qae} makes it crucial to find clean ways to extract it from experimental data. 
A common key assumption employed in the literature in order to be able to compare the CP asymmetry data to the SM at all is that the relative strong phase between CKM-subleading and CKM-leading amplitudes
\begin{equation}\label{eq:phase-def-1}
    \delta^f \equiv \arg \left(\frac{A_b^f}{A_d^f}\right)
\end{equation}
is $\mathcal{O}(1)$~\cite{Grossman:2019xcj}. It has been pointed out in the literature that strong phases corresponding to $\delta^f$ can be obtained via the measurements of time-dependent CP violation or quantum-correlated charm decays~\cite{Xing:1996pn, Gronau:2001nr, Bevan:2011up, Bevan:2013xla, Grossman:2012eb, Grossman:2019xcj, Xing:2019uzz, Schacht:2022kuj}. Additionally, it has been shown that the analogous strong phase in multi-body decays can be determined from a fit to the CP violating time-integrated Dalitz plot~\cite{Dery:2021mll}. 
However, even with recent experimental and methodological advances there exist no measurements yet of these strong phases, neither in two-body~\cite{LHCb:2021vmn, Pajero:2021jev}, nor in three-body~\cite{LHCb:2014nnj, LHCb:2023mwc, Davis:2023lxq, LHCb:2023rae} decays. 

In this letter, we assume the SM and show that isospin symmetry allows the determination of the relative strong phase between CKM-subleading and CKM-leading $D\rightarrow \pi\pi$ decays from direct CP asymmetries and branching ratios only. This enables at the same time also the extraction of the magnitude of the \lq\lq{}penguin over tree\rq\rq{} ratio $r^f$. 

The isospin construction proved very fruitful for the analysis of $B\rightarrow \pi\pi$ decays~\cite{Gronau:1990ka, Lipkin:1991st, Grossman:1997jr, Charles:1998qx, Gronau:2001ff, Charles:2004jd, Pivk:2004hq, Gronau:2005pq}.
Also in the case of $B$-decays it was found that the penguin over tree ratio as well as the relevant strong phase can be extracted from direct CP asymmetries and branching ratios, see, for example, Eqs. (7)--(9) and Eqs.~(110)--(114) in Ref.~\cite{Charles:1998qx} and the numerical results in Refs.~\cite{Charles:2006vd, Charles:2017evz}. 

Isospin symmetry has been applied to $D\rightarrow \pi\pi$ decays in Refs.~\cite{Grossman:2012eb, Franco:2012ck, Atwood:2012ac, Bevan:2013xla, Wang:2022nbm, Wang:2023pnb, Kwong:1993ri}. 
Note that the hierarchies of the interfering amplitudes in the $D$ and $B$ systems are very different from each other. Nevertheless, both systems have exactly the same group-theoretical structure under isospin, which implies that the two systems have the same sum rules at the amplitude level ~\cite{Gavrilova:2022hbx}. Thus we expect that similar isospin relations should hold at the level of observables as well. However, when deriving the implications for the observables, different approximations are used.

Below, after introducing our notation and approximations, we derive new isospin relations that allow the extraction of the ratio of CKM-subleading over CKM-leading amplitudes for $D^0\rightarrow \pi^+\pi^-$ and $D^0\rightarrow \pi^0\pi^0$. We study the numerical implications of current data and show the prospects of future more precise determinations of the penguin over tree ratio with future LHCb and Belle~II data. Afterwards, we conclude.

\section{Notation \label{sec:notation}}

Regarding Eq.~\eqref{eq:SCS-amps} we remark that we use a parametrization slightly different from the one frequently used for amplitudes of $D$-decays in the literature, see for example Ref.~\cite{Muller:2015rna}. In our case the CKM-leading contribution is accompanied by the CKM-factor $-\lambda_d$ and not $\lambda_{sd}\equiv (\lambda_s-\lambda_d)/2$. This choice ensures that the CKM-subleading part $A_b^f$ only contains contributions with isospin $\Delta I = 1/2$.
In general, the parametrization with the CKM-factor $\lambda_{sd}$ is a convenient choice when studying $U$-spin, while a parametrization with $\lambda_d$ is better suited for the isospin analysis of $D\rightarrow \pi\pi$. For brevity, we introduce the following notation for the decay amplitudes
\begin{align}
    A^{\pi^+\pi^-} &\equiv A^{+-}\,, &
    A^{\pi^0\pi^0} &\equiv A^{00}\,, &
    A^{\pi^+\pi^0} &\equiv A^{+0}\,.
\end{align}
For the relative strong phases of CKM-leading and CKM-subleading amplitudes we write 
\begin{align}\label{eq:strong_ph_def}
\delta^{+-} &\equiv \mathrm{arg}\left(\frac{A_b^{+-}}{A_d^{+-}}\right)\,, &
\delta^{00} &\equiv \mathrm{arg}\left(\frac{A_b^{00}}{A_d^{00}}\right)\,.
\end{align}
The phases $\delta^{+-}$ and $\delta^{00}$ enter the direct CP asymmetries of $D^0\rightarrow \pi^+\pi^-$ and $D^0\rightarrow \pi^0\pi^0$ decays, respectively.
We also use the relative strong phase between the CKM-leading contributions of the two decay channels, which we denote as
\begin{align}\label{eq:delta_d-def}
\delta_d &\equiv \mathrm{arg}\left(\frac{A_d^{+-}}{A_d^{00}}\right).
\end{align}
In the isospin limit, this phase is related to the relative phase between the $\Delta I=1/2$ and $\Delta I=3/2$ amplitudes which can be extracted from the $D\rightarrow \pi\pi$ branching ratios, see Ref.~\cite{CLEO:1993pky}.  
Direct CP asymmetries are defined as 
\begin{align}
{a}_{CP}^f &\equiv \frac{\vert A^f\vert^2 - \vert \overline{A}^f\vert^2}{ \vert A^f\vert^2 + \vert \overline{A}^f\vert^2 }\,.
\end{align}
The normalization of the amplitudes is such that the branching fractions are defined as
\begin{align}
\mathcal{B}^f = \mathcal{P}^f \cdot \frac{1}{2} \left( \vert A^f\vert^2 + \vert \bar{A}^f\vert^2 \right) \,,
\end{align}
where $\mathcal{P}^f$ are phase space factors given by
\begin{align}
\mathcal{P}^f &\equiv \frac{\tau_{D}}{16\pi m_D^3} \sqrt{ m_D^2 - ( m_{P_1} - m_{P_2})^2 }\, \times\nn\\
&\qquad\sqrt{m_D^2 - ( m_{P_1} + m_{P_2})^2}\,, \label{eq:phase-space}
\end{align}
where $D = D^0,\,D^\pm$ and $P_{1,2} = \pi^0,\,\pi^\pm$ depending on the specific final state $f$.

\section{Isospin Relations \label{sec:isospin-relations}}

\subsection{Approximations \label{sec:assumptions}}

Throughout this letter we consider the SM case only and use the following set of approximations.
\begin{enumerate}
    \item[$(i)$] \emph{We consider the isospin limit, i.e.,~we do not take into account isospin breaking effects.}
    \item[$(ii)$] \emph{We  neglect electromagnetic corrections and electroweak penguin contributions that are subleading due to the smallness of their Wilson coefficients.}
\end{enumerate}
The approximations $(i)$ and $(ii)$ are expected to hold at the $\mathcal{O}(1\%)$ level~\cite{Grossman:2012eb}.
Next, we use the fact that the involved CKM matrix elements are hierarchical 
and use the following approximations:
\begin{itemize}
\item[$(iii)$] \emph{For branching ratios, we neglect contributions which are suppressed by $\mathcal{O}(\vert\lambda_b/\lambda_d\vert)\sim \lambda^4$, where $\lambda\approx 0.23$ is the Wolfenstein parameter, such that}
\begin{align}\label{eq:B_CKM_lead}
\mathcal{B}^f = \mathcal{P}^f\cdot \abs{\lambda_d}^2\vert A^f_d\vert^2\,.
\end{align}
\item[$(iv)$]  \emph{We calculate direct CP asymmetries at leading order in} $\vert\lambda_b/\lambda_d\vert$~\cite{Golden:1989qx, Nierste:2017cua, Pirtskhalava:2011va} 
\begin{align}\label{eq:aCP_approx}
a_{CP}^f &= \mathrm{Im}\left(\frac{\lambda_b}{-\lambda_d}\right) r^f \sin(\delta^f)\,. 
\end{align}
\end{itemize}
Both Eqs.~(\ref{eq:B_CKM_lead}) and (\ref{eq:aCP_approx}) rely on $r^f$ respecting the hierarchy induced by the CKM matrix elements, that is, $r^f \ll 10^4$. Eq.~(\ref{eq:aCP_approx}) also relies on the convention that $A_d^f$ and $A_b^f$ contain strong phases only.

\subsection{Isospin decomposition \label{sec:isospin-decomposition}}

We write the isospin decomposition in such a way that all weak phases are explicit and the theory parameters depend on strong phases only~\cite{Schacht:2021jaz, Franco:2012ck, Atwood:2012ac}
\begin{align}
A^{+-} &= -\lambda_d \left( \sqrt{2} \left( t_{1/2} + t_{3/2}\right)\right) - \frac{\lambda_b}{2} \left( \sqrt{2} \, p_{1/2} \right) \,,\nonumber\\
A^{00} &= -\lambda_{d} \left( 2 t_{3/2} - t_{1/2}\right) -  \frac{\lambda_b}{2} \left(- p_{1/2} \right)\,, \nonumber\\
A^{+0}  &= -  \lambda_d \left( 3 \,t_{3/2}\right)\,. \label{eq:A-theory-param}
\end{align}
Here, $t_{1/2}$, $t_{3/2}$ are the CKM-leading contributions that correspond to $\Delta I=1/2$ and $\Delta I=3/2$ transitions, respectively, and $p_{1/2}$ is the CKM-subleading $\Delta I=1/2$ contribution. Note that assumptions $(i)$ and $(ii)$ ensure that in Eq.~(\ref{eq:A-theory-param}) there are no contributions with $\Delta I = 5/2$, and that there is no $\Delta I=3/2$ contribution to the CKM-subleading amplitude.
The notation in Eq.~(\ref{eq:A-theory-param}) is chosen such that one can directly read off $A_d^f$ and $A_b^f$ in the convention of Eq.~(\ref{eq:SCS-amps}). 

The isospin decomposition of the CP-conjugate amplitudes $\overline{A}^f$ takes the same form as Eq.~\eqref{eq:A-theory-param} up to the complex conjugation of the CKM-factors and unphysical overall phase-factors that do not enter any expressions for observables, see \emph{e.g.},~Ref.~\cite{Branco:1999fs}.

As the matrix elements $t_{1/2}$, $t_{3/2}$, and $p_{1/2}$ that enter the amplitudes $A^f$ and $\overline{A}^f$ are the same and we have more amplitudes than matrix elements, there exist sum rules between the amplitudes of the system~\cite{Grossman:2012eb}. As a consequence, this results also in relations between the observables that can be chosen as branching ratios and CP asymmetries. For example, from the last line of Eq.~(\ref{eq:A-theory-param}) it follows that $A^{+0} = \overline{A}^{-0}$, or in terms of observables, the well-known relation~\cite{Buccella:1992sg, Grossman:2012eb} 
\begin{align}
    a_{CP}^{+0} = 0. \label{eq:aCP+0-sum-rule}
\end{align}
Below we derive new isospin relations that relate branching fractions and CP asymmetries to $r^f$ and $\sin\delta^f$.
We then use these relations to extract the strong phases $\sin\delta^f$ and $r^f$
from direct CP asymmetries and branching ratio measurements only.

\subsection{Rescattering}
In principle, there exists an ambiguity in how one defines the two interfering amplitudes, \emph{e.g.}, we can define them with CKM-coefficients $-\lambda_d$ and $-\lambda_b/2$, as we do in Eq.~\eqref{eq:SCS-amps}, or for example with the CKM factors  $(\lambda_s-\lambda_d)/2$ and $-\lambda_b/2$, as frequently done when studying the $U$-spin system of neutral $D$ decays, see e.g.~Ref.~\cite{Grossman:2019xcj}.
Therefore the ``ratio of the two interfering amplitudes'' is not clearly defined a priori. 
An unambiguous definition can be obtained in the language of operator matrix elements.
Using the Hamiltonian and the notation of Ref.~\cite{Khodjamirian:2017zdu}
\begin{align}
\mathcal{H}_{\mathrm{eff}} &= \frac{G_F}{\sqrt{2}} \left( 
\sum_{q=d,s} \lambda_q \left(C_1 Q_1^q + C_2 Q_2^q\right)
\right) \equiv \sum_{q=d,s}\lambda_q\mathcal{O}^q\,,\\
Q_1^q &\equiv \left(\bar{u}\gamma_{\mu}(1-\gamma_5) q\right)  \left(\bar{q} \gamma_{\mu}(1-\gamma_5) c\right)\,, \\
Q_2^q &\equiv \left(\bar{q}\gamma_{\mu}(1-\gamma_5) q\right) 
\left(\bar{u} \gamma_{\mu}(1-\gamma_5) c\right)\,, \\
\langle \mathcal{O}^q\rangle^{f} &\equiv \bra{f} \mathcal{O}^q \ket{D^0}\,,
\end{align}
we write the decay amplitudes as 
\begin{align}
& A(D^0\rightarrow f) = \lambda_d \langle \mathcal{O}^d \rangle^f + \lambda_s \langle \mathcal{O}^s\rangle^f  \\
&\quad = -\lambda_d \left( \langle \mathcal{O}^s \rangle^f - \langle \mathcal{O}^d \rangle^f \right) 
- \frac{\lambda_b}{2}\left( 2 \langle \mathcal{O}^s \rangle^f \right)\,,
\end{align}
and define the penguin over tree ratio as
\begin{align}
r^f &\equiv \left|\frac{
2\langle \mathcal{O}^s \rangle^f}{
\langle \mathcal{O}^{s} \rangle^f  - \langle \mathcal{O}^{d} \rangle^f 
}\right|\,, \label{eq:definition-rf}
\end{align}
where we introduce the conventional factor of two in the numerator due to our convention Eq.~(\ref{eq:SCS-amps}). 
Once we fix the operator basis, our definition of $r^f$ becomes unambiguous.

We next elaborate on what we refer to as rescattering effects. In this letter we associate rescattering with the matrix element of the operator $\mathcal{O}^s$. The naive interpretation of rescattering in this case is that the $s \bar s$ pair rescatters into a $d \bar d$ pair that then hadronizes, together with the $u \bar u$ pair, into a pair of pions. Thus we define the no-rescattering limit as the limit in which
\beq
\langle \mathcal{O}^s\rangle_\text{no rescatt.} = 0\,,
\eeq
and consequently from Eq.~\eqref{eq:definition-rf}
\begin{equation}
    r^f_\text{no rescatt.} = 0\,. \label{eq:rf-norescattering}
\end{equation}

The limit of large rescattering on the other hand corresponds to the case $\vert\langle\mathcal{O}^s\rangle\vert \sim \vert\langle\mathcal{O}^d\rangle\vert$ which implies $r^f\gtrsim 1$, depending on the relative phase between $\langle O^d\rangle$ and $\langle O^s\rangle$. The experimental determination of $r^f$ thus provides a test of rescattering effects in $D\rightarrow \pi\pi$ decays.

We also mention one more useful limit, the $N_c \to \infty$ limit. A general discussion of the $1/N_c$ expansion in charm decays can be found in Ref.~\cite{Buras:1985xv}. This limit can be tested by studying a ratio of hadronic matrix elements~\cite{Buras:2014maa, Buras:1988ky} that can be defined as follows
\begin{equation}
r_{t} \equiv \abs{\frac{t_{1/2}}{t_{3/2}}}\,. \label{eq:large-Nc-tree}
\end{equation}
In the $N_c \rightarrow \infty$ limit one finds, completely analogous to the kaon sector~\cite{Buras:2014maa}, that
\begin{equation}
    r_{t}^{N_c\rightarrow \infty} = 2.
\end{equation}
In kaon physics, the fact that this ratio is much bigger than $2$ is referred to as the $\Delta I=1/2$ rule. The deviation of $r_t$ from $2$ can be interpreted as a failure of the $1/N_c$ expansion, sizable rescattering, significant kinematic effects or any combination of the above. The $r^f$ ratio, on the other hand, provides a clean probe of rescattering effects.

\subsection{Observables and Theory Parameters}

In our isospin analysis we use three types of parameters to which we refer as \emph{known parameters}, \emph{observables} and \emph{theory parameters}. 

We think of the CKM factors $\lambda_{d,b}$ and phase space factors $\mathcal{P}^f$ as known parameters. We assume that these parameters are well known from the independent experimental measurements and we use them as input in our analysis. 

The observables in our analysis are the branching fractions and CP asymmetries. For the system of $D \rightarrow \pi\pi$ decays, we thus have five observables
\begin{equation}\label{eq:exp-param}
    \mathcal{B}^{+-}, \quad \mathcal{B}^{00}, \quad \mathcal{B}^{+0}, \quad a_{CP}^{+-}, \quad a_{CP}^{00}\,.
\end{equation}

The theory parameters can be read off Eq.~\eqref{eq:A-theory-param} and are given by the magnitudes and relative phases of the hadronic matrix elements that appear on the RHS. In particular, without loss of generality, we choose $t_{3/2}$ to be real, resulting in a set of five theory parameters
\begin{equation}\label{eq:th-params}
    \abs{t_{3/2}}, \quad \abs{t_{1/2}}, \quad \abs{p_{1/2}}, \quad \arg \left(t_{1/2}\right), \quad \arg \left(p_{1/2}\right)\,.
\end{equation}
In what follows, we also use the following combinations of theory parameters in Eq.~\eqref{eq:th-params}
\begin{equation}\label{eq:th-params-more}
    r_t = \abs{\frac{t_{1/2}}{t_{3/2}}}, \quad \delta_t = \arg\left(\frac{t_{1/2}}{t_{3/2}}\right), \quad
    r^f, \quad \delta^f, \quad \delta_d\,,
\end{equation}
where $r^f$, $\delta^f$ and $\delta_d$ are defined in Eqs.~\eqref{eq:rf-def-1},~\eqref{eq:phase-def-1} and~\eqref{eq:delta_d-def}, respectively. We emphasize that the parameters in Eq.~\eqref{eq:th-params-more} are not independent from the parameters in Eq.~\eqref{eq:th-params} and among the two sets of the theory parameters there are only five that are independent.

The counting of the theory and experimental parameters shows that, in the isospin limit, the system of $D \rightarrow \pi\pi$ decays can be completely solved. That is, all the theory parameters in Eqs.~\eqref{eq:th-params}-\eqref{eq:th-params-more} can be expressed in terms of the observables listed in Eq.~\eqref{eq:exp-param}, possibly up to some discrete ambiguities. In particular, this means that the parameters of interest, $r^f$ and $\delta^f$, can be expressed in terms of branching fractions and direct CP asymmetries.

\subsection{Tree parameters}

Before we proceed to the determination of $r^f$ and $\sin\delta^f$, we elaborate on how to extract the phase $\delta_d$ 
defined in Eq.~\eqref{eq:delta_d-def} from the $D\rightarrow \pi\pi$ branching ratios in the isospin limit, see also Refs.~\cite{Franco:2012ck, CLEO:1993pky}.

From the isospin decomposition in Eq.~\eqref{eq:A-theory-param} and neglecting the CKM-subleading contributions to the amplitudes, 
we can solve for the theory parameters that contribute to the CKM-leading parts of the amplitudes, namely $\abs{t_{1/2}}$, $\abs{t_{3/2}}$ and the relative phase $\delta_t$ between $t_{1/2}$ and $t_{3/2}$. We obtain 
\begin{align}
\vert t_{3/2}\vert &= \frac{1}{3\abs{\lambda_d}}\sqrt{\frac{\mathcal{B}^{+0}}{\mathcal{P}^{+0}}}\,,\label{eq:abs_t32}\\
\vert t_{1/2}\vert &= \frac{1}{\vert \lambda_d\vert} \sqrt{\left( 
	\frac{1}{3} \frac{\mathcal{B}^{+-}}{\mathcal{P}^{+-}} +
	\frac{1}{3} \frac{\mathcal{B}^{00}}{\mathcal{P}^{00}} -
	\frac{2}{9} \frac{\mathcal{B}^{+0}}{\mathcal{P}^{+0}}
	\right)}\,, \\
r_{t} &\equiv \frac{\vert t_{1/2}\vert}{\vert t_{3/2}\vert} = 
\sqrt{3 \frac{ \mathcal{B}^{+-}}{ \mathcal{B}^{+0}  } \frac{\mathcal{P}^{+0}}{\mathcal{P}^{+-}}  +
3 \frac{ \mathcal{B}^{00} }{ \mathcal{B}^{+0}} \frac{\mathcal{P}^{+0}}{\mathcal{P}^{00}} -2
}\,, \label{eq:br-abs} \\
\cos\delta_{t}&\equiv \cos\left(\mathrm{arg}\left(\frac{t_{1/2}}{t_{3/2}}\right)\right) \nonumber \\
&= \frac{1}{2 r_{t}} \left( 
    \frac{9}{2} \frac{\mathcal{B}^{+-}}{ \mathcal{B}^{+0}} \frac{\mathcal{P}^{+0}}{\mathcal{P}^{+-}} -
    r_{t}^2  - 1
    \right)
	\,. \label{eq:br-phase}
\end{align}
From these relations and Eq.~\eqref{eq:A-theory-param} we derive the expression for the phase $\delta_d$ between the CKM-leading amplitudes of $D\rightarrow \pi^+\pi^-$ and $D\rightarrow \pi^0\pi^0$ as defined in Eq.~\eqref{eq:delta_d-def}. We have
\begin{align}
   \cos \delta_d &=\frac{2- r_{t}^2 + r_{t}  \cos \delta_{t}}{\sqrt{\left(2-r_{t}^2 + r_{t} \cos \delta_{t}\right)^2+9 r_{t}^2  \sin ^2\delta_{t}}}\,,
    \end{align}
and
\begin{align}\label{eq:sin_dd}
    \sin \delta_d &= \frac{\pm 3 r_{t} \vert\sin \delta_{t}\vert}{\sqrt{\left(2-r_{t}^2+ r_{t} \cos \delta_{t}\right)^2+9 r_{t}^2 \sin^2\delta_{t}}}\,.
\end{align}
Note that $\sin\delta_d$ is only known up to a sign, as we do not know the sign of $\sin\delta_{t}$.
Our results Eqs.~(\ref{eq:br-abs}) and (\ref{eq:br-phase}) agree with the analytical expressions in Ref.~\cite{CLEO:1993pky}, once we account for the different conventions of the isospin decomposition.

A comment is in order about the phase $\delta_d$. The phase $\delta_d$ is defined as a relative phase between CKM-leading amplitudes of $D^0 \rightarrow \pi^+\pi^-$ and $D^0 \rightarrow \pi^0 \pi^0$, two decays with different final states. In general, such a relative phase can not be physical, yet here we solve for $\delta_d$ in terms of physical observables. The resolution of this apparent paradox is that in the isospin limit $\pi^{\pm}$ and $\pi^0$ are not distinguishable, which makes this phase meaningful in the isospin limit.

\subsection{Penguin over Tree Ratio}

We employ now the isospin amplitude sum rule~\cite{Grossman:2012eb, Grossman:2012ry, Atwood:2012ac} for the CKM-suppressed amplitudes. 
From Eq.~\eqref{eq:SCS-amps} and the first and second lines of Eq.~\eqref{eq:A-theory-param} it follows that
\begin{align}
    \frac{1}{\sqrt{2}} A_b^{+-} = -A_b^{00}\,,\label{eq:isospin-Ab}
\end{align}
which implies the following relations for the absolute values and the relative phase of CKM-subleading amplitudes:
\begin{equation}\label{eq:Abpm/Ab00_abs}
    \abs{\frac{A_b^{+-}}{A_b^{00}}} = \sqrt{2}\,,
\end{equation}
\begin{equation}\label{eq:Abpm/Ab00_arg}
    \arg\left(\frac{A_b^{+-}}{A_b^{00}}\right) = \pi\,.
\end{equation}
We note that Eqs.~(\ref{eq:isospin-Ab})--(\ref{eq:Abpm/Ab00_arg}) are related to the amplitude sum rule in Eq.~(14) of Ref.~\cite{Grossman:2012eb}.

In the following we derive the implications of Eqs.~\eqref{eq:Abpm/Ab00_abs} and~\eqref{eq:Abpm/Ab00_arg}. We obtain two isospin limit relations that relate branching fractions, CP asymmetries and the relative strong phases $\delta^{+-}$ and $\delta^{00}$ defined in Eq.~\eqref{eq:strong_ph_def}. We then use these two relations to solve for the relative strong phases in terms of branching fractions and CP asymmetries. Finally, we use the approximation in Eq.~\eqref{eq:aCP_approx} to solve for the penguin over tree ratio.

First, we derive the implications of the sum rule between the absolute values of CKM-subleading amplitudes given in Eq.~\eqref{eq:Abpm/Ab00_abs}. 
Dividing the expressions Eq.~(\ref{eq:aCP_approx}) for CP asymmetries, we obtain
\begin{align}\label{eq:obs_SR}
\frac{ \sin\delta^{+-}}{\sin\delta^{00} } &= \frac{ a_{CP}^{+-} }{ a_{CP}^{00}  }\sqrt{\frac{1}{2} \frac{ \mathcal{B}^{+-}}{\mathcal{P}^{+-}}  \frac{\mathcal{P}^{00}  }{\mathcal{B}^{00} }} \,,
\end{align}
and 
\begin{align}
\frac{r^{00}}{r^{+-}}  &= \sqrt{\frac{1}{2}\frac{ \mathcal{B}^{+-}}{\mathcal{P}^{+-}}  \frac{\mathcal{P}^{00}  }{\mathcal{B}^{00} }}\,. \label{eq:penguin-ratio-ratio}
\end{align}
We emphasize that this relation only holds when all of the approximations~(\emph{i})-(\emph{iv}) are satisfied.

Next, we derive the implication of the sum rule between the arguments of the CKM-subleading amplitudes given in Eq.~\eqref{eq:Abpm/Ab00_arg}.
We write $\delta^{+-}$ as 
\begin{align}\label{eq:phase_SR}
\delta^{+-} &= \mathrm{arg}\left( \frac{A_b^{+-}}{A_d^{+-}}\right)
	     = \mathrm{arg}\left( \frac{A_b^{+-} A_d^{00} A_b^{00} }{A_d^{+-} A_d^{00} A_b^{00} } \right) \\
      	     &= \pi + \delta^{00} -\delta_d\,,
\end{align}
where in the last line we used the isospin relation in Eq.~\eqref{eq:Abpm/Ab00_arg}.
This result allows us to express the LHS of Eq.~\eqref{eq:obs_SR} in terms of the phases $\delta^{00}$ and $\delta_d$. We find
\begin{align}\label{eq:sin/sin}
\frac{\sin\delta^{+-}}{\sin\delta^{00}} &= \cot\delta^{00} \sin \delta_d -\cos\delta_d \,.
\end{align}
Now, as we have two relations that relate branching ratios, CP asymmetries and relative strong phases, Eqs.~\eqref{eq:obs_SR} and~\eqref{eq:phase_SR}, we can solve for $\sin \delta^{+-}$ and $\sin \delta^{00}$. To do so, we substitute the result of Eq.~\eqref{eq:sin/sin} into Eq.~\eqref{eq:obs_SR} and solve for $\cot\delta^{00}$. We find
\begin{align}
\cot\delta^{00} &= \frac{1}{\sin\delta_d}\frac{ a_{CP}^{+-} }{ a_{CP}^{00}  }\sqrt{\frac{1}{2}\frac{ \mathcal{B}^{+-}}{\mathcal{P}^{+-}} \frac{\mathcal{P}^{00}  }{ \mathcal{B}^{00} } } + \cot\delta_d\,,
\end{align}
from which we can obtain
\begin{align}
&\sin\delta^{00} = \frac{-\text{sign\,}(a_{CP}^{00})}{\sqrt{1 + \frac{1}{ \sin^2\delta_d  } \left( \frac{ a_{CP}^{+-} }{ a_{CP}^{00}  } \sqrt{\frac{1}{2}\frac{ \mathcal{B}^{+-}}{\mathcal{P}^{+-}}  \frac{\mathcal{P}^{00}  }{ \mathcal{B}^{00} }} + \cos \delta_d \right)^2  }}\,. \label{eq:phase00}
\end{align}
Similarly we obtain 
\begin{align}\label{eq:phase+-}
    \sin \delta^{+-} &= \frac{-\text{sign\,}(a_{CP}^{+-})}{\sqrt{1 + \frac{1}{\sin^2\delta_d}\left(\frac{a_{CP}^{00}}{a_{CP}^{+-}}\sqrt{2\frac{\mathcal{P}^{+-}}{\mathcal{B}^{+-}}\frac{\mathcal{B}^{00}}{\mathcal{P}^{00}}} +\cos\delta_d\right)^2}}\,.
\end{align}
The sign of each of the $\sin \delta^{f}$ can be extracted from the sign of $a_{CP}^f$ using the fact 
that $\text{Im\,}\left(-\lambda_b/\lambda_d\right) < 0$.
The ratio of the CKM-subleading over CKM-leading amplitudes follows then from Eq.~\eqref{eq:aCP_approx}
\begin{align}\label{eq:Abf/Adf}
 r^f  &= \left| \frac{ a_{CP}^f }{ \sin\delta^f \, \mathrm{Im}\left(-\lambda_b/\lambda_d\right)}\right| \,.	
\end{align}
Substituting the expressions for $\sin\delta^f$ in Eqs.~\eqref{eq:phase00} and~\eqref{eq:phase+-} into Eq.~\eqref{eq:Abf/Adf}, we arrive at
\begin{align}
 & r^{00} = \frac{1}{\vert \mathrm{Im}\left(-\lambda_b/\lambda_d\right)\vert}\times\nn\\&
\sqrt{
    (a_{CP}^{00})^2 +
    \frac{(a_{CP}^{+-} \sqrt{\mathcal{B}^{+-} \mathcal{P}^{00}} + a_{CP}^{00} \sqrt{2\mathcal{B}^{00}\mathcal{P}^{+-}} \cos\delta_d )^2}{
    2 \mathcal{B}^{00} \mathcal{P}^{+-} \sin^2\delta_d }
    }\,, \label{eq:PT00} 
\end{align}
and 
\begin{align}\label{eq:rpm}
    &r^{+-} = \frac{1}{\abs{\text{Im}\left(-\lambda_b/\lambda_d\right)}} \times\nn\\& \sqrt{\left(a_{CP}^{+-}\right)^2 + \frac{\left(a_{CP}^{00}\sqrt{2\mathcal{B}^{00}\mathcal{P}^{+-}} + a_{CP}^{+-}\sqrt{\mathcal{B}^{+-}\mathcal{P}^{00}}\cos \delta_d\right)^2}{\mathcal{B}^{+-} \mathcal{P}^{00} \sin^2\delta_d}}\,,
\end{align}
where $\sin \delta_d$ and $\cos \delta_d$ are given in Eqs.~\eqref{eq:br-abs}-\eqref{eq:sin_dd} in terms of branching fractions only. Eqs.~\eqref{eq:phase00}-\eqref{eq:rpm} are the main results of this letter. They allow the extraction of $r^f$ and $\sin \delta^f$ from direct CP asymmetries and branching fractions with a theory uncertainty of $\mathcal{O}(1\%)$.

\section{Numerical Results \label{sec:numerics}}

\begin{table}[t]
\centering
\begin{tabular}{|c|c|c|}
\hline\hline 
\multicolumn{3}{|c|}{Direct CP Asymmetries} \Tstrut\\[1pt]
\hline
$a_{CP}^{+0}$ &  $+0.004 \pm 0.008$ \Tstrut\Tstrut & \cite{HeavyFlavorAveragingGroup:2022wzx,LHCb:2021rou, Belle:2017tho, CLEO:2009fiz} 
\\[2pt]
$a_{CP}^{00}$ &  $-0.0002\pm 0.0064$ & $^a$\cite{HeavyFlavorAveragingGroup:2022wzx, Belle:2014evd, CLEO:2000opx}~ \\[2pt]
$a_{CP}^{+-}$ & $0.00232 \pm 0.00061$  & \cite{LHCb:2022lry} \\[2pt]
\hline
\multicolumn{3}{|c|}{Branching Ratios} \Tstrut\Tstrut \\[2pt]
\hline
$\mathcal{B}(D^0\rightarrow \pi^+\pi^0)$ \Tstrut    & $(1.247\pm 0.033)\cdot 10^{-3}$  & \cite{Workman:2022ynf}\\[2pt]

~$\mathcal{B}(D^0\rightarrow\pi^+\pi^-)$ ~& ~$(1.454 \pm 0.024)\cdot 10^{-3}$  ~ & \cite{Workman:2022ynf}\\[2pt]

$\mathcal{B}(D^0\rightarrow \pi^0\pi^0)$  & $( 8.26 \pm 0.25 )\cdot 10^{-4}$  & \cite{Workman:2022ynf}\\[2pt]
\hline
\multicolumn{3}{|c|}{Further Numerical Inputs}\Tstrut  \Tstrut  \\[2pt]
\hline
$\mathrm{Im}\left(\lambda_b/(-\lambda_d)\right)$ \Tstrut  & $(-6.1 \pm 0.3)\cdot 10^{-4}$  & \cite{Workman:2022ynf} \\[2pt]
\hline\hline
\end{tabular}
\caption{Experimental input data.  
We use the decay times and masses from Ref.~\cite{Workman:2022ynf}. $^a$Our extraction from $A_{CP}(D^0\rightarrow \pi^0\pi^0) = -0.0003\pm 0.0064$~\cite{HeavyFlavorAveragingGroup:2022wzx} and $\Delta Y = (-1.0\pm 1.1\pm 0.3)\cdot 10^{-4}$~\cite{LHCb:2021vmn}. 
\label{tab:input}}
\end{table}

\begin{table}[t]
\centering
\begin{tabular}{|c|c|}
\hline\hline \Tstrut  \Tstrut  
~$a_{CP}^{+-}$~ & ~$(2.32\pm 0.07)\cdot 10^{-3}$~  \\[2pt]
$a_{CP}^{00}$ &  $(-2 \pm 9)\cdot 10^{-4}$\\[2pt]\hline\hline 
\end{tabular}
\caption{Future data scenario employing the current central values and using prospects for the errors from 
Table 6.5~of Ref.~\cite{LHCb:2018roe}~(300~fb$^{-1}$) and Table~122~of Ref.~\cite{Belle-II:2018jsg}~(50 ab$^{-1}$) for $D^0\rightarrow \pi^+\pi^-$ and $D^0\rightarrow \pi^0\pi^0$, respectively. All other input data is left as specified in Table~\ref{tab:input}.
\label{tab:future-data}}
\end{table}

\begin{table}[t]
\centering
\begin{tabular}{|c|c|c|}
\hline\hline
~Parameter~ & ~Current data~ & ~Future data scenario~ \Tstrut  \Tstrut  \\[2pt]
\hline
$r_{t}$ \Tstrut & $3.43 \pm 0.06$ & $3.43 \pm 0.06$\\[2pt]
$\cos\delta_{t}$ & $0.06 \pm 0.02$ & $0.06 \pm 0.02$\\[2pt]
$\cos\delta_d$ & $-0.68\pm 0.01$  & $-0.68\pm 0.01$ \\[2pt]
$\vert \sin\delta^{00}\vert$ &  $0^{+1}_{-0}$ & $0.06^{+0.20}_{-0.06}$\\[2pt]
$\vert \sin\delta^{+-}\vert$ &  $0.7^{+0.3}_{-0.5}$ & $0.69^{+0.21}_{-0.16}$ \\[2pt]
$r^{00}$ &  $5.2^{+13.3}_{-2.4}$ & $5.2^{+1.6}_{-1.2}$  \\[2pt]
$r^{+-}$ & $5.5^{+14.2}_{-2.7}$ & $5.5^{+1.8}_{-1.3}$ \\[2pt]
\hline\hline
\end{tabular}
\caption{Numerical results for current and hypothetical future data. 
In the future data scenario, the results for $r_t$, $\cos\delta_{t}$ and $\cos\delta_d$ are identical to the ones with current data, as these depend only on the branching ratio data which is not modified in the future data scenario compared to current data. Furthermore, in the future data scenario $\sin\delta^{+-}<0$. The overall additional relative systematic uncertainty of $\mathcal{O}(10\%)$ due to the universality assumption of $\Delta Y$ for the extraction of the direct CP asymmetries comes on top of the errors shown here, see text for details.
\label{tab:results}}
\end{table}

\begin{table}[t]
\centering
\begin{tabular}{|c|c|}
\hline\hline
 ~Hypothesis~ & ~Current data~ \Tstrut\Tstrut \\ \hline 
$r^{+-} = 1.0$ \Tstrut &  $2.7\sigma$ \\[2pt]
$r^{+-} = 0.1$ & $3.7\sigma$ \\[2pt]
$r^{00} = 1.0$ & $2.6\sigma$ \\[2pt]
$r^{00} = 0.1$ & $3.7\sigma$ \\[2pt]
$p_{1/2} = 0$ & $3.8\sigma$ \\[2pt]\hline\hline
\end{tabular}
\caption{Test of benchmark hypotheses and significance of their rejection for current data. 
In the considered future data scenario all hypotheses listed here are rejected at $>5\sigma$.
In order to account for the overall $\mathcal{O}(10\%)$ relative systematic uncertainty due to the assumption of a universal $\Delta Y$ for the extraction of the direct CP asymmetries, we multiply the hypotheses for $r^f$ by a factor $1.10$, resulting in a more conservative (lower) significance of rejection, see text for details. 
\label{tab:tests}}
\end{table}

In the following we show numerical results based on the formalism presented above.
We use the numerical input as given in Table~\ref{tab:input}.
We also consider a future data scenario which is specified in Table~\ref{tab:future-data}.
Note that the value of $a_{CP}^{+-}$ that we use has been extracted by LHCb employing a universal time-dependent CP violation coefficient $\Delta Y$~\cite{LHCb:2022lry, Gersabeck:2011xj} and we use the same value for $\Delta Y$ also for our extraction of $a_{CP}^{00}$ from the corresponding time-integrated measurement, see Table~\ref{tab:input}. A universal $\Delta Y$ is motivated by $U$-spin symmetry~\cite{Kagan:2020vri}. This implies an overall systematic theory uncertainty in the extraction of the direct CP asymmetries of second order in $U$-spin breaking, see Eq.~(133)~in~Ref.~\cite{Kagan:2020vri}. This uncertainty can be generically estimated as~$\mathcal{O}(10\%)$. 
Thus in order for the numerical predictions for $\delta^f$ and $r^f$ to reach a theory uncertainty of $\mathcal{O}(1\%)$, improved measurements that do not require the universality assumption for $\Delta Y$ are necessary. This can be achieved by employing future data on $\Delta Y^{+-}$ and $\Delta Y^{00}$ in the determination of the direct CP asymmetries. We emphasize that in the following we use the data on $a_{CP}^{+-}$ and $a_{CP}^{00}$ with the universality assumption in place. This results consequently in an overall theory uncertainty of roughly $\mathcal{O}(10\%)$ on top of the experimental errors quoted below for the numerical values of $\sin\delta^f$ and $r^f$ as obtained from  Eqs.~\eqref{eq:phase00}-\eqref{eq:rpm}.

Our results are given in Table~\ref{tab:results}. 
We also perform tests of benchmark scenarios for $r^f$ and list their significance of rejection with current data in Table \ref{tab:tests}.

We make the following observations:
\begin{itemize}
\item The measurement of $a_{CP}^{+0}$ agrees with the isospin sum rule Eq.~(\ref{eq:aCP+0-sum-rule}).
\item Although we have currently essentially no information about $\sin\delta^{00}$, we can still infer non-trivial information about $r^{00}$. This can be understood from closer inspection of Eq.~(\ref{eq:PT00}), which contains a contribution independent of $a_{CP}^{00}$. This is a consequence of isospin symmetry, which translates knowledge about $D^0\rightarrow \pi^+\pi^-$ into constraints on $D^0\rightarrow \pi^0\pi^0$, see also Eq.~(\ref{eq:penguin-ratio-ratio}).
\item The central values of $r^{00}$ ($r^{+-}$) are quite large and $2.6\sigma$ ($2.7\sigma$) away from 1.
\item The future data scenario demonstrates that our method will allow to extract the ratio of CKM-subleading over CKM-leading amplitudes with unprecedented precision, the key advantage being that no assumption about the strong phases $\delta^{00}$ and $\delta^{+-}$ has to be made in order to do so.
\item Once the sign of $a_{CP}^f$ is determined, we also know the sign of $\sin\delta^f$. In our future data scenario this is the case for $\sin\delta^{+-}$.
\item Comparing $r_t$ to the limit Eq.~(\ref{eq:large-Nc-tree}), we observe an $\mathcal{O}(1)$ enhancement in the \lq\lq{}charm $\Delta I=1/2$ rule\rq\rq{}, in agreement with Ref.~\cite{Grossman:2019xcj}.
\item Comparing to the limit of no rescattering Eq.~(\ref{eq:rf-norescattering}) we obtain for current data $r^f > 0$ at $3.8\sigma$.
\end{itemize}
The last observation, if confirmed, implies either very large rescattering effects or BSM physics.

\section{Conclusion \label{sec:conclusions}}

We show that the approximate isospin symmetry of the $D\rightarrow \pi\pi$ system allows for the  extraction of the magnitudes of CKM-subleading over CKM-leading amplitudes from direct CP asymmetries and branching ratios without making assumptions about the relevant strong phase.
This ratio is colloquially known as the \lq\lq{}penguin over tree ratio\rq\rq{}. 
Our main theoretical results are given in Eqs.~\eqref{eq:phase00}-\eqref{eq:rpm}. Numerically, from current data we obtain
\begin{align}
r^{00} &= 5.2^{+13.3}_{-2.4}\,,\\
r^{+-} &= 5.5^{+14.2}_{-2.7}\,.
\end{align}
The large central values are driven by the sizable direct CP asymmetry found currently in $D\rightarrow \pi^+\pi^-$. 
Using prospects for the development of experimental errors, we demonstrate that our method will allow to probe the puzzle of penguin enhancement in charm decays with unprecedented precision in the future.

\section{Acknowledgments}

We thank Marco Gersabeck and Ulrich Nierste for useful discussions. 
The work of MG was supported in part by the National Science Foundation under Grant No. NSF PHY-1748958 and NSF PHY-2309135, the Heising-Simons Foundation, and the Simons Foundation (216179, LB).
YG is supported in
part by the NSF grant PHY-2014071. S.S. is supported by a Stephen Hawking Fellowship from UKRI under reference EP/T01623X/1 and the STFC research grants ST/T001038/1 and ST/X00077X/1. For the purpose of open access, the authors have applied a Creative Commons Attribution (CC BY) licence to any Author Accepted Manuscript version arising. This work uses existing data which is available at locations cited in the bibliography. 

\bibliography{draft.bib}
\bibliographystyle{apsrev4-1}

\end{document}